\documentclass{article}
\usepackage{amsmath}
\usepackage{arxiv}

\usepackage[utf8]{inputenc} 
\usepackage[T1]{fontenc}    
\usepackage{hyperref}       
\usepackage{url}            
\usepackage{booktabs}       
\usepackage{amsfonts}       
\usepackage{nicefrac}       
\usepackage{microtype}      
\usepackage{cleveref}       
\usepackage{lipsum}         
\usepackage{graphicx}
\usepackage{natbib}
\usepackage{doi}
\usepackage{subcaption}
\title{Enhancing Sentinel-2 Image Resolution: Evaluating Advanced Techniques based on Convolutional and Generative Neural Networks}

\newif\ifuniqueAffiliation
\uniqueAffiliationtrue

\ifuniqueAffiliation 
\author{{Patrick Kramer}\thanks{This work has been submitted to the IEEE for possible publication. Copyright may be transferred without notice, after which this version may no longer be accessible.} \\
	Fraunhofer Austria Research GmbH\\
	9020 Klagenfurt am Worthersee, Austria \\
	\texttt{patrick.kramer@fraunhofer.at} \\
	\And
        Coauthor \\
	{Alexander Steinhardt} \\
	Fraunhofer Austria Research GmbH\\
	9020 Klagenfurt am Worthersee, Austria \\
	\texttt{alexander.steinhardt@fraunhofer.at} \\
        \And
        Coauthor \\
	{Barbara Pedretscher} \\
	Fraunhofer Austria Research GmbH\\
	9020 Klagenfurt am Worthersee, Austria \\
	\texttt{barbara.pedretscher@fraunhofer.at} \\
}
\else
\usepackage{authblk}

\newbox{\orcid}\sbox{\orcid}{\includegraphics[scale=0.06]{orcid.pdf}} 

\fi

\renewcommand{\shorttitle}{ ENHANCING SENTINEL-2 IMAGE RESOLUTION }

\begin{document}
\maketitle

\begin{abstract}
This paper investigates the enhancement of spatial resolution in Sentinel-2 bands that contain spectral information using advanced super-resolution techniques by a factor of 2. State-of-the-art CNN models are compared with enhanced GAN approaches in terms of quality and feasibility. Therefore, a representative dataset comprising Sentinel-2 low-resolution images and corresponding high-resolution aerial orthophotos is required. Literature study reveals no feasible dataset for the land type of interest (forests), for which reason an adequate dataset had to be generated in addition, accounting for accurate alignment and image source optimization. The results reveal that while CNN-based approaches produce satisfactory outcomes, they tend to yield blurry images. In contrast, GAN-based models not only provide clear and detailed images, but also demonstrate superior performance in terms of quantitative assessment, underlying the potential of the framework beyond the specific land type investigated.
\end{abstract}


\section{Introduction}
Satellite images in remote sensing are indispensable for numerous applications that monitor changes in the atmosphere and on the earth surface, both on land and at sea. Many of these applications make use of the data products of the Copernicus Sentinel-2 mission of the European Space Agency providing  13 multi-spectral bands ranging from the visible to the short-wave infrared spectrum \citep{Drusch2012}. The resolution of the observation data enables a wide range of monitoring applications. However, the available resolution limits show severe restrictions when it comes to the precise detection of fine details and small objects. Currently,  commercial satellites or aerial images with a correspondingly higher resolution are used as a workaround to overcome that specific drawback.  In general, these products are of elevated costs, and thus not suitable for large-scale time-series analyses \citep{Yoo2021}. To overcome and therewith enhance the advantages of the Sentinel-2 mission, such as the open data policy, five-day revisit frequency, and global coverage, towards increased public applications the possibility of technological resolution boosting would be beneficial. For that reason, Super-Resolution (SR) techniques are investigated in terms of their resolution upgrade capabilities. More precisely, representative tuples of high-resolution (HR) orthographic images and low-resolution (LR) Sentinel-2 data are combined to learn and develop an AI model that synthetically upgrades the resolution for the defined land type of interest. The model is learned over a representative area of interest (AoI), which in the following can serve as baseline for a meta model incorporating heterogeneous geographical structures. Therefore, a representative dataset out of publicly available aerial orthophotos with a spatial resolution of \(20 \times 20\) cm cm serving as ground truth (GT) data and correspondingly Sentinel-2 data, which are 10 meters in resolution, is formed. The aim is to increase the spatial resolution of the Sentinel-2 RGB bands by a factor 2.  In this paper different SR approaches are investigated and compared in terms of their resulting image quality.  This comparison is performed using state-of-the-art pixel-based metrics such as the Peak Signal-to-Noise Ratio (PSNR) and the Structural Similarity Index Measure (SSIM) quantifying image quality. In addition, the Learned Perceptual Image Patch Similarity (LPIPS) metric is evaluated, as it mimics human perception \citep{Zhang2018} and thus enables a more comprehensive assessment of the obtained image quality.
\section{Related Work and Methodological Foundations}\label{RelatedWork}
In the field of SISR, where, as the name indicates, a single LR image of a scenery is used to construct an HR output, interpolation algorithms, such as bilinear, bicubic and spline interpolations, are widely applied because of their simple mathematical operations, which made them popular in early studies \citep{Karwowska2022}, \citep{Kirkland2010}, \citep{Keys1981}. However, exactly their simplicity limits their ability to enhance image quality especially when complex structures are present. In general, they tend to blur the images as they cannot sufficiently reconstruct high-frequency details that are crucial for image sharpness. \citep{Lanaras2018}.

A different approach in this context is pan-sharpening, which involves merging an HR panchromatic (PAN) image with an LR multi-spectral image to create an HR super-resolved image. The fused product incorporates both the high spatial resolution of the PAN image and the rich multi spectral information of the LR image. Although some of these methods have been successfully applied, cf. \citep{Vivone2015}, \citep{Loncan2015} and \citep{Zhou2020},  they are not directly applicable to Sentinel-2, as no PAN band is available for that mission.
\begin{table*}[!hb]
    \caption{Overview on Spectral and Spatial Characteristics of S2 Bands\citep{Spoto2012}.\label{tab:sentinel2bands}}
    \centering
    \renewcommand{\arraystretch}{1.2}
    \begin{tabular}{lccc}\hline
    \textbf{Band} & \textbf{Central Wavelength (nm)} & \textbf{Bandwidth (nm)} & \textbf{Spatial Resolution-GSD (m)} \\
    \hline
    B01: Coastal Aerosol & 443 & 20  & 60 \\
    B02: Blue            & 490 & 65  & 10 \\
    B03: Green           & 560 & 35  & 10 \\
    B04: Red             & 665 & 30  & 10 \\
    B05: Red-edge 1      & 705 & 15  & 20 \\
    B06: Red-edge 2      & 740 & 15  & 20 \\
    B07: Red-edge 3      & 783 & 20  & 20 \\
    B08: Near-IR         & 842 & 115 & 10 \\
    B08A: Near-IR Narrow& 865 & 20  & 20 \\
    B09: Water Vapor     & 945 & 20  & 60 \\
    B10: SWIR Cirrus    & 1375 & 20 & 60 \\
    B11: SWIR           & 1610 & 90 & 20 \\
    B12: SWIR           & 2190 & 180 & 20 \\
    \hline
    \end{tabular}
\end{table*}

The latest and most innovative approach in the field of remote sensing resolution boosting is based on deep learning models -- an approach that appears promising for Sentinel-2 and thus forms the baseline for this work. A significant milestone in the SR domain was achieved with the Super-Resolution Convolutional Neural Network (SRCNN), presented in \cite{Dong2016}. The architecture of this model consists of multiple convolutional layers that extract and reconstruct fine details from the LR image. Each layer performs a specific function, ranging from feature extraction to nonlinear mapping and reconstruction, to ensure that the final product is not only higher in resolution but also richer in details and textures. In \cite{Liebel2016}, Liebel and Körner demonstrated the effectiveness of SRCNN for satellite imagery, particularly for Sentinel-2 images, where they first upscale the LR image to the size of the HR counterpart by bicubic interpolation before applying the SRCNN network. In \cite{Fu2018}, Fu et al. present an alternative SISR method based on a Remote Sensing CNN (RSCNN) that builds on the basic structure of SRCNN. This model is characterized by multiple convolutional layers that add depth to the network, effectively countering blurring effects. Quantitatively, RSCNN shows better PSNR values compared to standard SRCNN. Further, it provides better visual results, highlighting its effectiveness in enhancing remote sensing images. In the SRCNN architecture the LR image is initially upscaled to the size of the HR image using bicubic interpolation before being fed into the network. To minimize computational power, the Fast Super-Resolution Convolutional Neural Network (FSRCNN) was developed, which performs enhancement directly on the LR image before upsampling to the required HR output size \cite{Dong2016a}. For performance and quality tests, Kawulok et al. created a dataset out of Sentinel-2 HR images, which were downsampled using various interpolation methods to generate LR images \cite{Kawulok2019}.  The FSRCNN was compared with a significantly more complex network, the SRResNet -- a very deep residual network presented in \cite{Ledig2017}. This architecture consists of residual blocks, each containing two convolutional layers followed by batch normalization (BN), and upsampling blocks that enlarge the image and thus improve the resolution. However, Kawulok et al. point out that SRResNet does not necessarily outperform the comparatively simpler FSRCNN on the tested dataset. Authors like Galar et al. apply Enhanced Deep Residual Networks (EDRS), which use modified SRResNet models \cite{Galar2019}.

Important to mention at this point are the approaches that are based on Generative Adversarial Networks (GANs). Basically a GAN model consists of two networks: a generator and a discriminator \cite{Goodfellow2020}. One of the first GAN architectures in the field of SISR was introduced by Ledig et al., who used the SRResNet as a generator architecture and a perceptual loss function to generate photo-realistic natural images \cite{Ledig2017}. This SRGAN framework was successfully applied on Sentinel-2 images, where PeruSat-1 images formed the HR reference. The results remarkably demonstrated the effectiveness of the approach \cite{Pineda2020}. The Enhanced Super-Resolution Generative Adversarial Network (ESRGAN) was developed to minimize the artifacts resulting from the SRGAN framework and to generate visually better images \cite{Wang2018}. Here, the residual blocks of SRResNet were replaced with Residual-in-Residual Dense Blocks (RRDB), eliminating the computationally intensive BN. Additionally, the discriminator was revised: Rather than assessing the authenticity of a singular image, the relativistic methodology quantifies the comparative realism between two images. Salgueiro Romero et al. show that the ESRGAN architecture works with pairs of WorldView-Sentinel images, aiming to generate a super-resolved multi-spectral Sentinel-2 output \cite{SalgueiroRomero2020}. In the field of remote sensing, the Enlighten-GAN presented by Gong et al. in \cite{Gong2021} represents a further development of the ESRGAN. This architecture ensures that the network, when trained with satellite images, converges to a reliable point. To achieve performance improvement, a Self-Supervised Hierarchical Perceptual Loss is used. Moreover, the large-scale images are divided into smaller patches and then reassembled, resulting in more efficient memory utilization. An improvement over ESRGAN is the Real-ESRGAN, presented by Wang et al. in \cite{Wang2021}. The objective of this architecture is to enhance the resolution and quality of LR images that suffer from unknown and complex degradation, commonly found in real-world scenarios. Fundamentally, Real-ESRGAN differs from ESRGAN primarily in the discriminator component, which employs a U-Net architecture with spectral normalization. This architecture has outperformed previously described models on synthetic datasets. However, a comparison on a real-world use case, specifically improving Sentinel-2 images, has not been performed yet. Therefore, the following sections will focus on the application and comparison of these models to enhance the resolution of Sentinel-2 images.

\section{Materials and Methods}
\label{sec:materialsmethods}
\subsection{Dataset generation}
\label{sec:Dataset}
In literature, two prominent datasets for SR in the field of remote sensing exist: the WorldStrat \cite{Cornebise2022} and the SEN2VENµS \cite{Michel2022} datasets. Whereas the first can not be used because of present cloudy patches, the second one does not provide images that are processed with the standard SEN2Cor \cite{Louis2016} pre-processing algorithm, which is essential for a subsequent SR methodology. For these reasons, an adequate dataset had to be created. As the focus of the SR empowering application of interest is related to forest land cover, tuples of S2 data and aerial orthophotos from representative regions of Carinthia, Austria, were used. At this point it is important to mention that the Sentinel-2 mission comprises two polar-orbiting satellites, S2A and S2B, which are offset by 180° to each other, see \cite{Drusch2012}. This constellation enables a high revisit frequency of 5 days at the equator, which allows precise and continuous monitoring of changes on the earth's surface. Altogether, the mission includes 13 multi-spectral bands, ranging from visible and near-infrared to shortwave infrared, with different spatial resolutions. There are four bands with \(10 \times 10\) m, six bands with \(20 \times 20\) m and three bands with \(60 \times 60\) m resolution \cite{Spoto2012}. Table~\ref{tab:sentinel2bands} summaries the spectral and spatial characteristics of Sentinel-2, a distinction is made between two different product types: Level-1C, which represents the top-of-atmosphere reflectance (TOA), and Level-2A, which represents the bottom-of-atmosphere reflectance (BOA). Level-1C includes radiometric and geometric corrections. In contrast, Level-2A provides atmospherically corrected surface reflectance images derived from the associated Level-1C products. For the Level-2A correction, the above mentioned Sen2Cor algorithm \cite{Louis2016} is typically used, accounting for the scattering of air molecules, the absorbing and scattering effects of atmospheric gases, especially ozone, oxygen and water vapor, and the absorption and scattering by aerosol particles \cite{Drusch2012,Spoto2012}. For the application and comparison of different SR methods in this study, Level-2A products are primarily used as LR data source, with a specific focus on the three RGB bands. For the HR counterpart, digital aerial orthophotos from the state of Carinthia are used. These data are publicly available via the KAGIS portal\footnote{\url{http://kagis.ktn.gv.at}} \cite{KAGIS} and comprise four spectral bands (RGB and NIR) with a spatial resolution of \(20 \times 20\) cm.

\begin{figure}[!ht]
\centering
\includegraphics[width=3.5in]{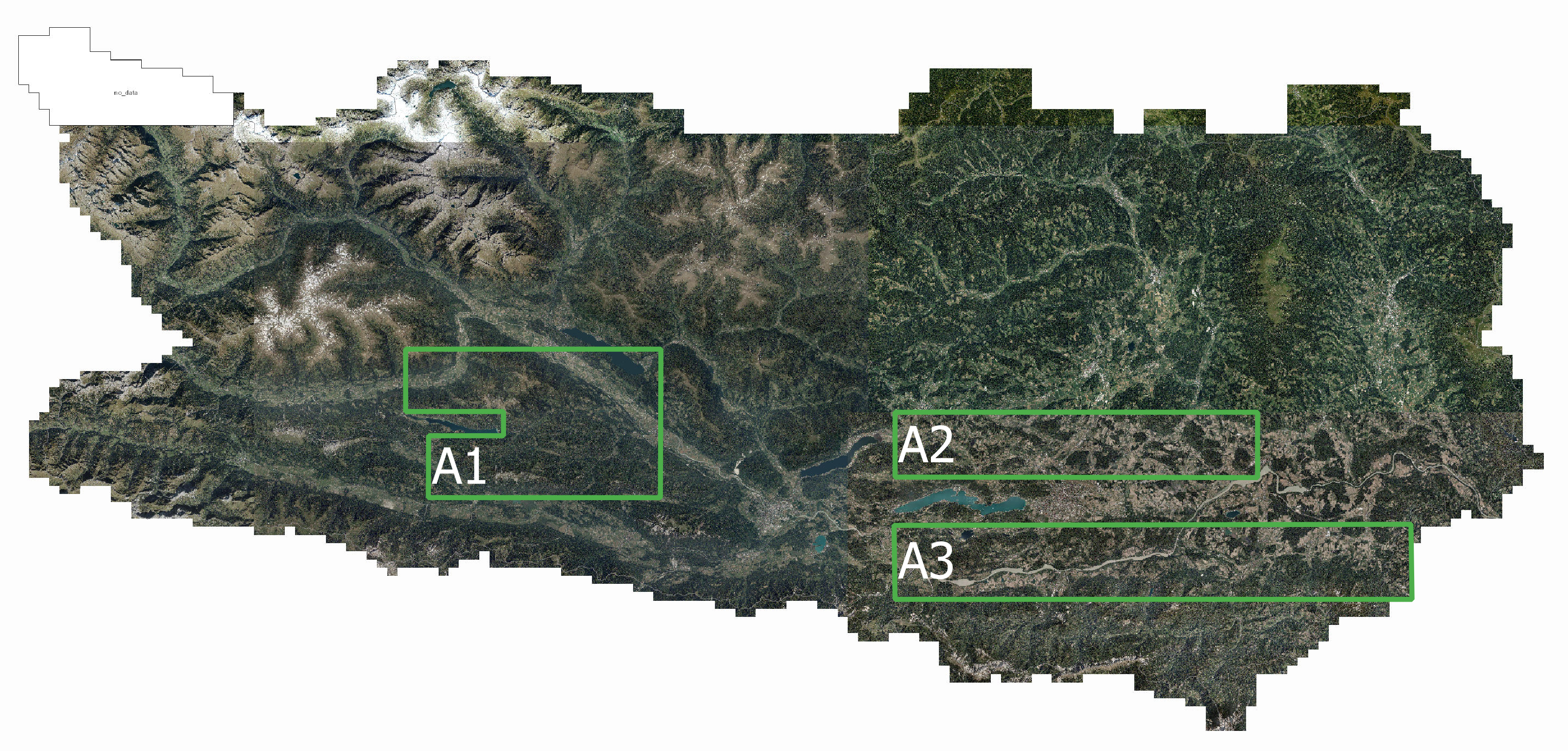}
\caption{Aerial orthophotos of Carinthia as provided by \cite{KAGIS}, highlighting the areas of interest.}
\label{fig_1}
\end{figure}

\begin{figure}[ht!]
\centering
\begin{subfigure}[b]{\linewidth}
    \centering
    \includegraphics[width=0.55\linewidth]{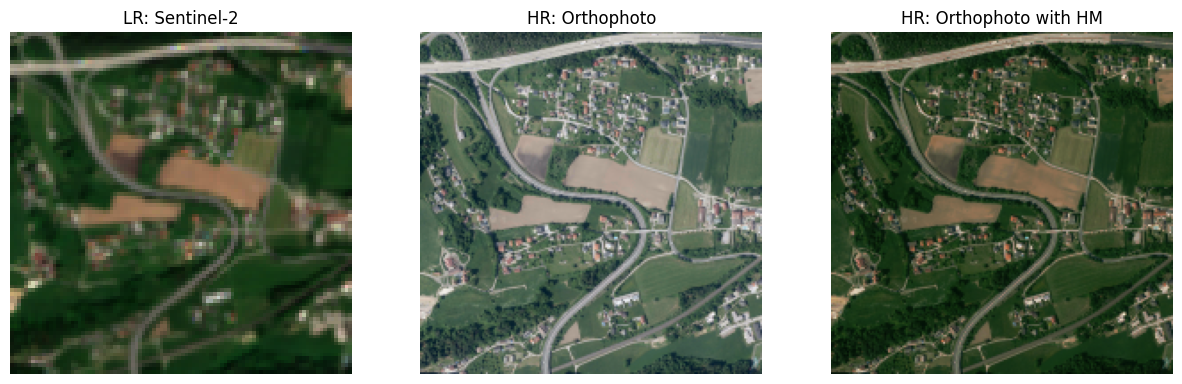}
    \vspace{10pt}
    \centering
    \includegraphics[width=0.55\linewidth]{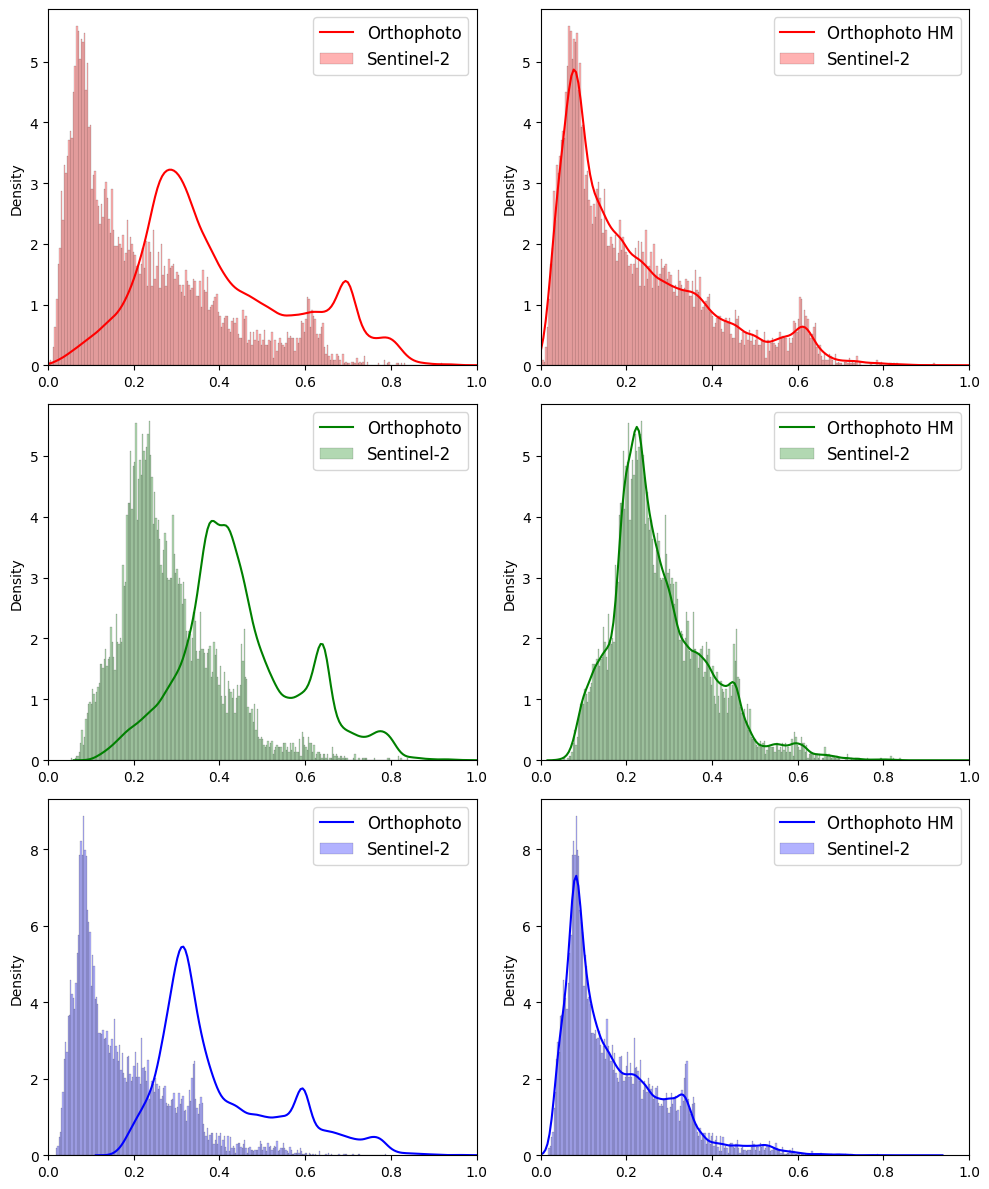}
\end{subfigure}
\caption{Visual representation of the histogram matching process.}
\label{fig:histmatching}
\end{figure}

For the generation of a valid dataset, it is crucial that both image sources are optimal registered in time, coupled with minimal cloud interference. The chosen aerial orthophotos of Carinthia were recorded between 2018 and 2022. Accordingly, suitable Sentinel-2 images with less than 5 \% cloud cover have to be found for every GT tile. This pre-filtering boundary condition led to the defined AoIs, shown in Figure \ref{fig_1} with details given in Table~\ref{tab:ortho_s2}.
\begin{table}[ht!]
    \caption{Comparative data: Aerial orthophotos and Sentinel-2 data.}\label{tab:ortho_s2}
    \centering
    \renewcommand{\arraystretch}{1.2}
    \begin{tabular}{ccccc}
    \hline
    \textbf{AoI} & \textbf{Year} & \textbf{Aerial orthophoto} & \textbf{S2} & \textbf{Area (km\textsuperscript{2})} \\
    \hline
    A1 & 2022 & 20.06 & 19.06 & $\sim$ 900 \\
    A2 & 2022 & 19.05 & 20.05 & $\sim$ 700 \\
    A3 & 2022 & 19.05 & 20.05 & $\sim$ 1000 \\
    \hline
    \end{tabular}
\end{table}Given the AoIs, next both image sources had to be converted into the same coordinate reference system (CRS). Therefore, the aerial orthophoto, which is available in the Austrian coordinate system EPSG:31255, is projected onto the CRS of the S2 data, which is WGS 84 / EPSG:32633. An aerial orthophoto typically covers an area of about 1.25 x 1 km. As both images are now available in the same CRS, this area is cut out of the larger 110 x 110 km Sentinel-2 tile, resulting in matching sections. These sections are then used for further processing and data generation. After careful temporal alignment, which had to be performed by visual comparison due to the fact that suitable Sentinel-2 images are not available for every single date, and cropping of corresponding image sections, additional pre-processing steps were carried out. A significant difference between the two image sources lies in the fact that the images are captured with two different sensor types, which leads to differences in bit depth. Aerial orthophotos are recorded with a camera that provides natural images with a sampling depth of 8 bits per pixel. In contrast, sensors for multi-spectral remote sensing produce images with a sampling depth of 12 bits per pixel \cite{Drusch2012}. For data consistency, both image sources need to be normalized first with respect to their bit depth. Therefore, a min-max normalization is chosen guaranteeing values within the compact interval \([0, 1] \). Next, the high-frequency components of HR aerial orthophotos are reduced by means of arithmetic mean filtering to minimize aliasing effects and reduce noise. Therefore, an \(n \times n\) box kernel is used, where \(n\) is calculated based on the spatial resolution of the aerial orthophoto and the desired output resolution of \(5 \times 5\) m. For the setting under investigation, this results in a \(25 \times 25\) box kernel, where during filtering, the average value is calculated. After noise reduction, the HR image is downscaled to the desired resolution of \(5 \times 5\) m using bicubic interpolation. This method enables a smooth and accurate representation of the image data by calculating a better approximation of the pixel values between the known pixel positions \cite{Gonzalez2017}. The final pre-processing step involves spectral adjustment, which is performed by histogram matching, as explained in \cite{Gonzalez2017}. In principle this method aims to match the histograms of two image sources based on the underlying cumulative distribution functions. Mapping is achieved by adjusting the pixel value in the target image based on the corresponding value in the reference. Representative results of the histogram adjustment for the three RGB bands are shown in Figure \ref{fig:histmatching}.

After data pre-processing, patches from the Sentinel-2 data as well as from the bicubic interpolated aerial orthophotos are extracted. For the LR images, these patches have a size of \(96 \times 96\) px with a spatial resolution of \(10 \times 10\) m, while the HR patches have a size of \(192 \times 192\) px with a spatial resolution of \(5 \times 5\) m. Finally, to remove pairs of non-matching information due to e.g. cloud coverage and thus to obtain a valid dataset for training quality filtering based on the SSIM and PSNR metrics is performed, whereby lower than 0.45 (21) in SSIM (PSNR) are removed. Table~\ref{tab:dataset_distribution} summaries the main statistics for the dataset.
\begin{table}[ht!]
    \centering
    \caption{Dataset splitting for SR model development}
    \label{tab:dataset_distribution}
    \renewcommand{\arraystretch}{1.2}
    \begin{tabular}{lccc}
    \hline
    \textbf{Dataset} & \textbf{Number of patches} & \textbf{LR patch size} & \textbf{HR patch size} \\ \hline
    Train            & 1500 & \(96 \times 96\) px & \(192 \times 192\) px                     \\ 
    Validation       & 374 & \(96 \times 96\) px & \(192 \times 192\) px                     \\ 
    Test             & 208 & \(96 \times 96\) px  & \(192 \times 192\) px                   \\ 
    \hline
    \end{tabular}
\end{table}

\subsection{Super Resolution Methods}
\label{sec:SRMethods}
\begin{figure*}[ht!]
\begin{center}
		\includegraphics[width=\columnwidth]{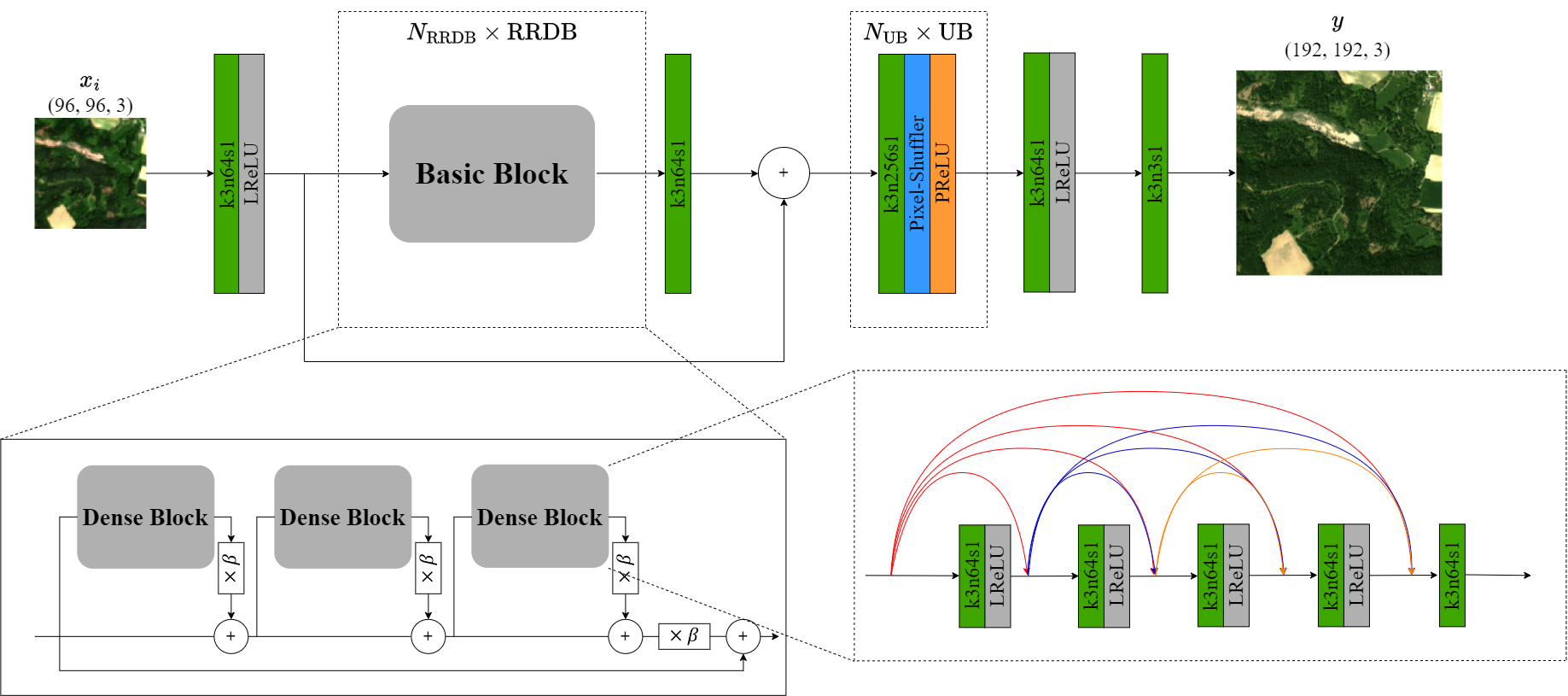}
	\caption{Generator architecture described in \cite{Wang2018} with \( N_{\text{RRDB}} = 4 \) and \( N_{\text{UB}} = 1 \).}
\label{fig:esrgangen}
\end{center}

\end{figure*}

\begin{figure}[ht!]
\centering
\begin{subfigure}{\linewidth}
    \centering
    \includegraphics[width=\linewidth]{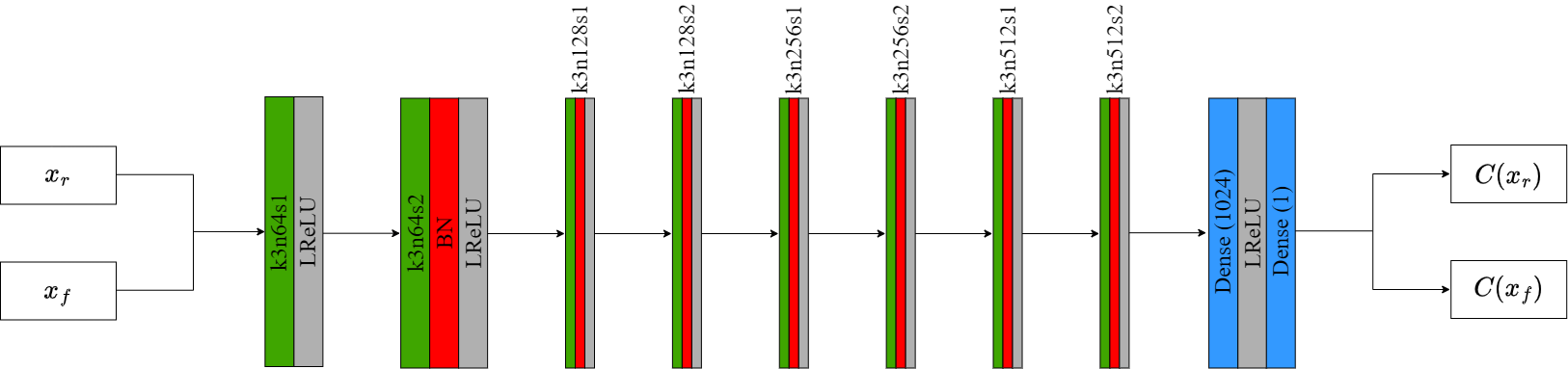}
    \subcaption{}
    \vspace{10pt}
    \label{fig:esrgan_disc}
    \centering
    \includegraphics[width=\linewidth]{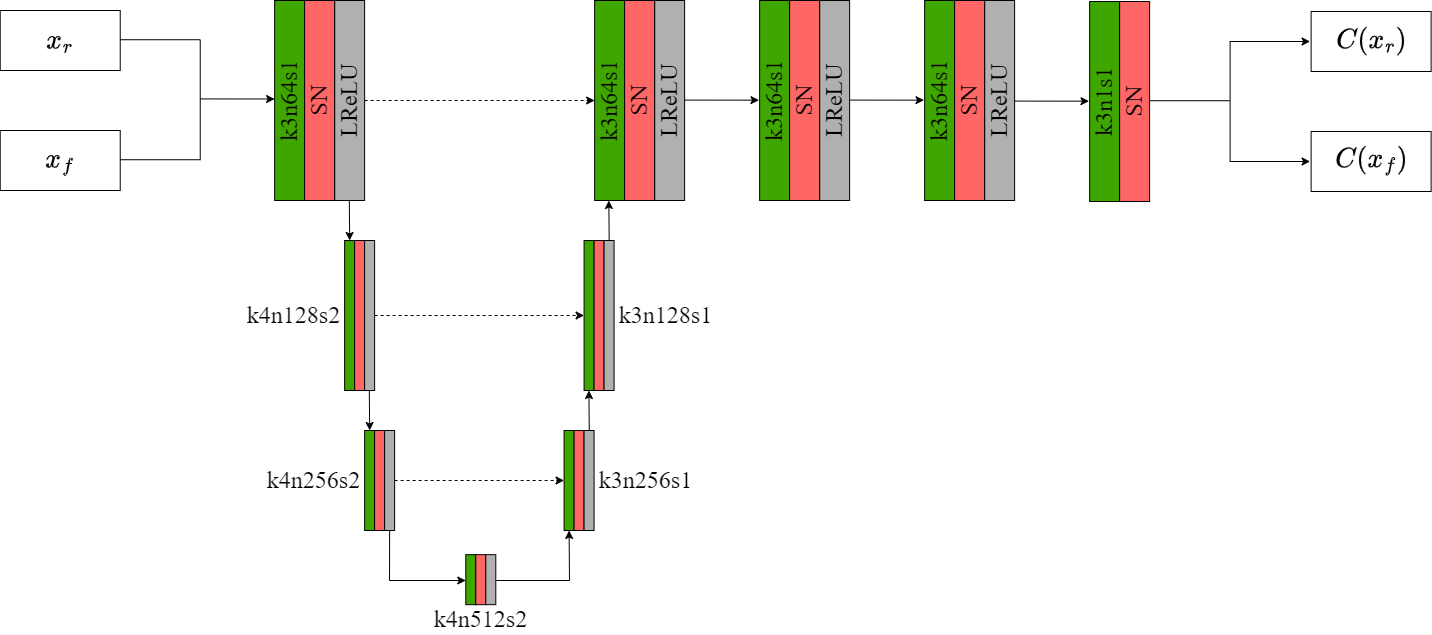}
    \subcaption{}
    \label{fig:real_esrgan_unet_disc}
\end{subfigure}
\caption{Discriminator Architectures: (a) ESRGAN proposed \cite{Wang2018} in with classic blocks, (b) Real-ESRGAN proposed \cite{Wang2021} with U-Net structure.}
\label{fig:esrgan_discriminators}
\end{figure}

For the scope of this work, two different CNN-based approaches are first examined as a baseline: the SRCNN \cite{Dong2016} and the deeper residual network SRResNet \cite{Ledig2017}. Furthermore, as an alternative to the CNN-based approaches, the ESRGAN \cite{Wang2018} and Real-ESRGAN \cite{Wang2021} architectures are investigated in terms of their up-sampling capabilities. As mentioned in Chapter~\ref{RelatedWork}, a GAN consists of a generator and a discriminator, which are trained simultaneously through an adversarial process. The generator has the task of generating data, while the discriminator acts as a probabilistic classifier and assesses whether a generated image should be classified as real or fake \cite{Goodfellow2020}. Thus, the goal of the generator is to maximize the probability that the discriminator fails increasingly improving the quality of the generated data and therewith producing realistic images that cannot be identified as fakes anymore. This adversarial process results in high-quality super-resolved images that are rich in detail. The generator architecture for both, ESRGAN and Real-ESRGAN, is based on the work of Wang et al. \cite{Wang2018}. Figure~\ref{fig:esrgangen} schematically illustrates its structure with the corresponding configurations for the kernel sizes (k), number of feature maps (n), and strides (s) for every convolutional layer. The fundamental element is built upon the Residual-in-Residual Dense Blocks (RRDB), which combine multi-level residual networks and dense connections without a BN layer. Another key component is the up-sampling block (UB), which uses a pixel shuffler to transfer values from the channel dimension to the height-width dimension, thereby enlarging the image. The number of RRDBs and UBs is variable -- a higher number of RRDBs makes the network more complex and enables better mapping, whereas increasing the number of UBs results in a higher scaling factor. In this work, \( N_{\text{RRDB}} = 4 \) and \( N_{\text{UB}} = 2 \) are used to upscale from a spatial resolution of \(10 \times 10\) m to \(5 \times 5\) m.
The main difference between the above mentioned GAN architectures lies in the discriminator network. The network used in ESRGAN is a classical classification network consisting of eight blocks, each comprising a convolution layer followed by BN and a leaky ReLU layer (LReLU). The final layers are two dense fully-connected layers that combine the extracted features to an one-dimensional vector. In contrast, Real-ESRGAN employs a U-Net discriminator with spectral normalization (SN) layers to enhance training stability and reduce the oversharp and disturbing artifacts introduced by the GAN training. Moreover, the U-Net architecture ensures that, instead of discriminating against global styles, accurate gradient feedback is provided for local textures. The structures of both discriminator architectures and their corresponding parametrization in terms of kernel, feature maps and strides, can be seen in Figure~\ref{fig:esrgan_discriminators}.
The training and optimization process is divided into two phases, following the framework provided in \cite{Wang2018}. Initially, the generator is pre-trained to avoid undesired local optima using the expected \(L_{1}\) loss
\begin{equation}\label{equ:2}
    L_{1} = \mathbb{E}_{x_i} \| G(x_i) - y \|_1,
\end{equation}
where \(L_{1}\) defines the mean absolute error between the generated image \(G(x_i)\), with \(x_i\) defining the \(i^{\text{th}}\) input image, and the ground truth \(y\). Next, the pre-trained generator is optimized together with the discriminator following an adversarial approach, where the weights of the generator are updated according to the total loss of the generator:
\begin{equation}\label{equ:3}
    L_G = L_{\text{percep}} + \lambda L_G^{Ra} + \eta L_1.
\end{equation}As can be seen, the total generator loss \(L_G\) is a combination of perceptual loss \(L_{\text{percep}}\), adversarial loss \(L_G^{Ra}\), and content loss \(L_{1}\), with the coefficients \(\lambda\) and \(\eta\) balancing the contribution of the individual loss terms. 
For better interpretation, note that the perceptual loss function, as introduced in \cite{Johnson2016}, aims to better adapt to the human sense of perception. For this purpose, a pre-trained 19-layer VGG network \cite{Simonyan2015} is utilized, and the mean absolute error between the feature maps -- obtained after the activation function of the \(4\)-th convolutional layer and before the \(5\)-th max-pooling layer -- of \(G(x_i)\) and \(y\) is calculated. 
The relativistic average discriminator approach is calculated according to the method presented in \cite{Wang2018}. It assesses whether a real input image is comparatively more realistic than a fake one, which is formulated as \( D_{Ra}(x_r, x_f) = \sigma(C(x_r)) - \mathbb{E}_{x_f}[C(x_f)] \), where \( \sigma(\cdot) \) is the sigmoid function, \( C(x) \) represents the output of the discriminator, and \( \mathbb{E}_{x_f}(\cdot) \) is the average over all fake data in the mini-batch. Subsequently, \(L_G^{Ra} \)is calculated using cross-entropy, as shown in Equation~\ref{equ:4}, where \( x_f = G(x_i) \) and \( x_r = y \).
\begin{equation}\label{equ:4}
\begin{split}
L_G^{Ra} = & -\mathbb{E}_{x_r}\left[\log(1 - D_{Ra}(x_r, x_f))\right] \\
           & -\mathbb{E}_{x_f}\left[\log(D_{Ra}(x_r, x_f))\right]
\end{split}
\end{equation}
In GAN framework, which optimize a min-max problem, the discriminator loss can be calculated in a symmetrical form as follows:
\begin{equation}\label{equ:5}
\begin{split}
    L_D^{Ra} = & -\mathbb{E}_{x_r}\left[\log(D_{Ra}(x_r, x_f))\right] \\
               & -\mathbb{E}_{x_f}\left[\log(1 - D_{Ra}(x_r, x_f))\right]    
\end{split}
\end{equation}
\subsection{Image Quality Assessment}\label{sec:IQA}
In this work we use PSNR, SSIM, and LPIPS metrics to assess the quality of the SR output. PSNR is a pixel-based metric, expressed in decibels, that represents the ratio between the maximum intensity value present in an image and the noise. PSNR is calculated according to
\begin{equation}\label{equ:6}
    PSNR = 10 \cdot \log_{10} \left( \frac{\text{MAX}^2}{\text{MSE}} \right),
\end{equation}where MSE refers to the Mean Squared Error and MAX to the maximum pixel intensity value of the image under investigation. The second pixel-based metric in the above list, SSIM, see e.g. \cite{Wang2004}, measures the similarity between two images by comparing windows of size \(N \times N\) comprising structure information, luminance and contrast. In contrast to these pixel-metrics, the LPIPS metric however showed to be more suitable for the application of interest as it reflects human perception \cite{Zhang2018}. For computation, images pass a pre-trained deep neural network such as AlexNet, VGG or ResNet. The activation's, i.e. outputs at certain layers, are then extracted, and used to quantitatively assess the differences in terms of MSE between two images.
\section{Experiments and Results}
\label{sec:results}
\begin{figure*}[ht!]
\centering
\begin{subfigure}[b]{\linewidth}
    \centering
    \includegraphics[width=\linewidth]{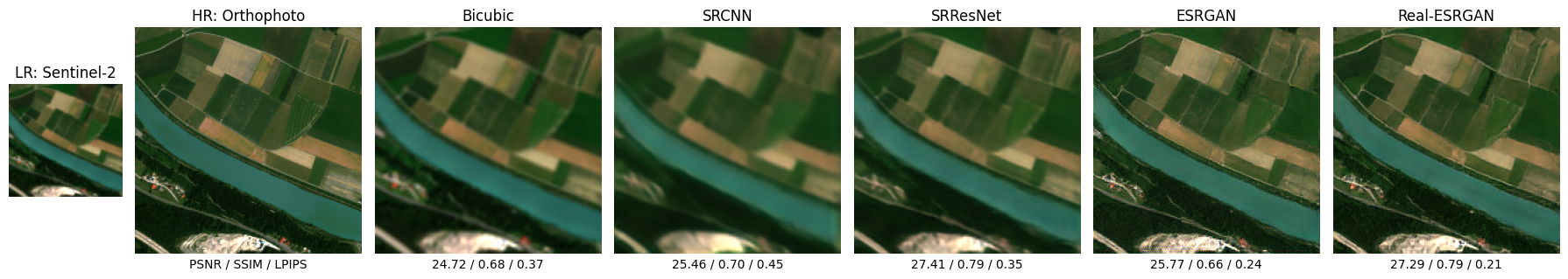}
\end{subfigure}
\begin{subfigure}[b]{\linewidth}
    \centering
    \includegraphics[width=\linewidth]{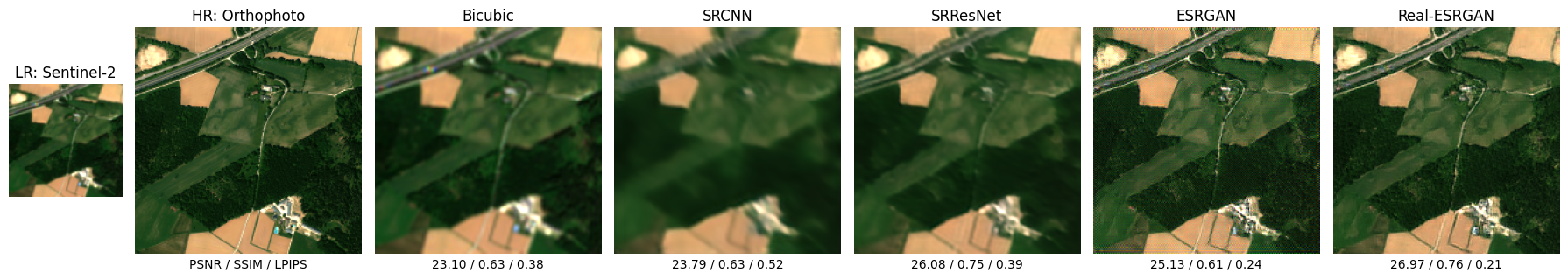}
\end{subfigure}
\begin{subfigure}[b]{\linewidth}
    \centering
    \includegraphics[width=\linewidth]{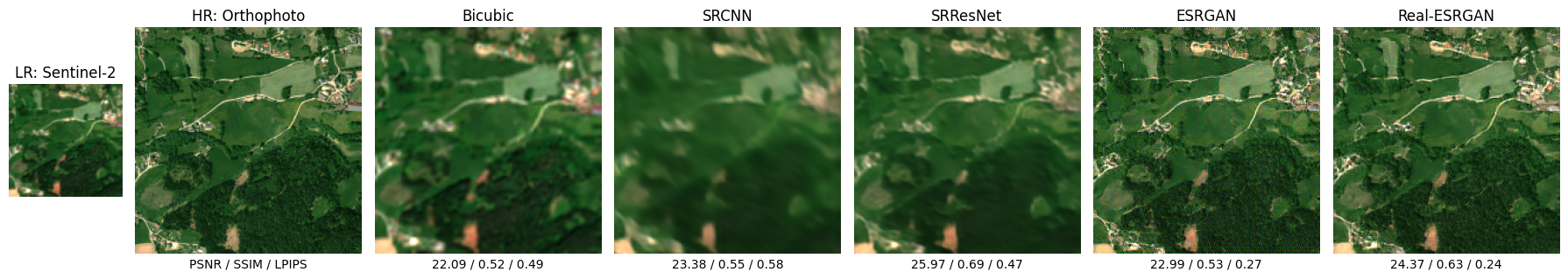}
\end{subfigure}
\begin{subfigure}[b]{\linewidth}
    \centering
    \includegraphics[width=\linewidth]{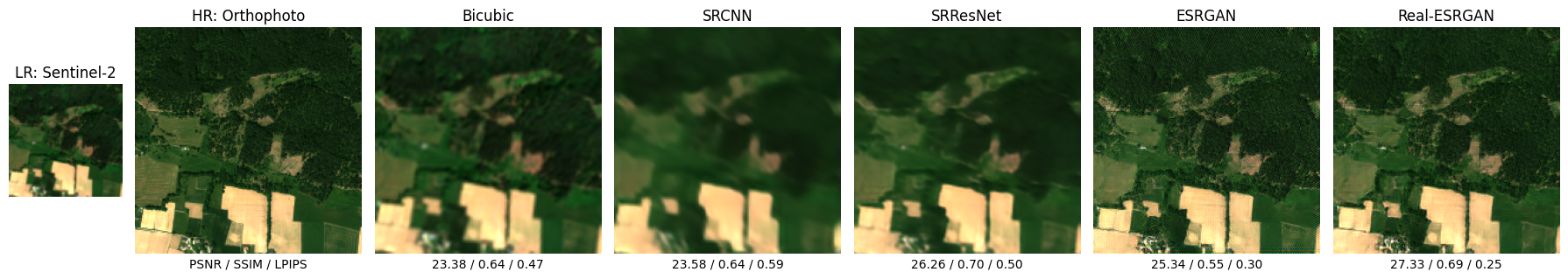}
\end{subfigure}
\begin{subfigure}[b]{\linewidth}
    \centering
    \includegraphics[width=\linewidth]{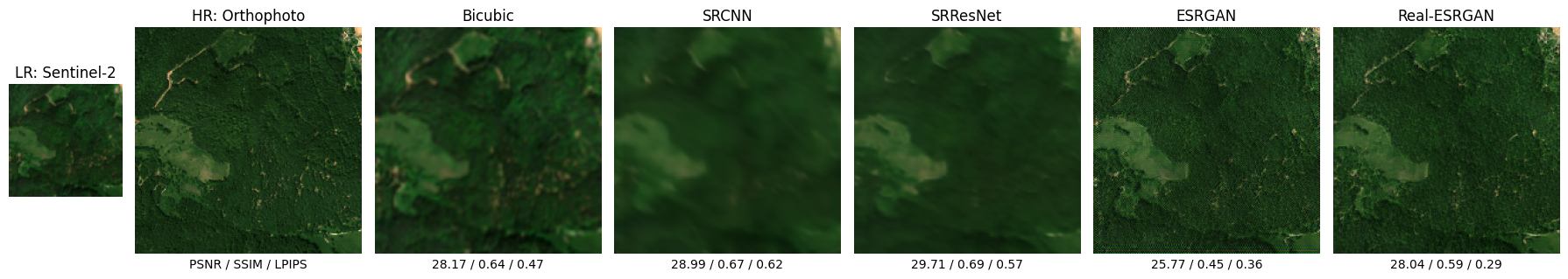}
\end{subfigure}
\begin{subfigure}[b]{\linewidth}
    \centering
    \includegraphics[width=\linewidth]{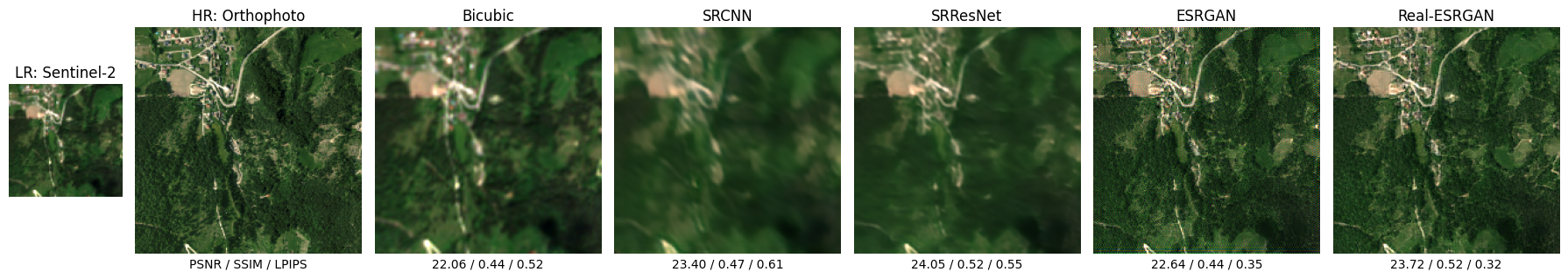}
\end{subfigure}
\begin{subfigure}[b]{\linewidth}
    \centering
    \includegraphics[width=\linewidth]{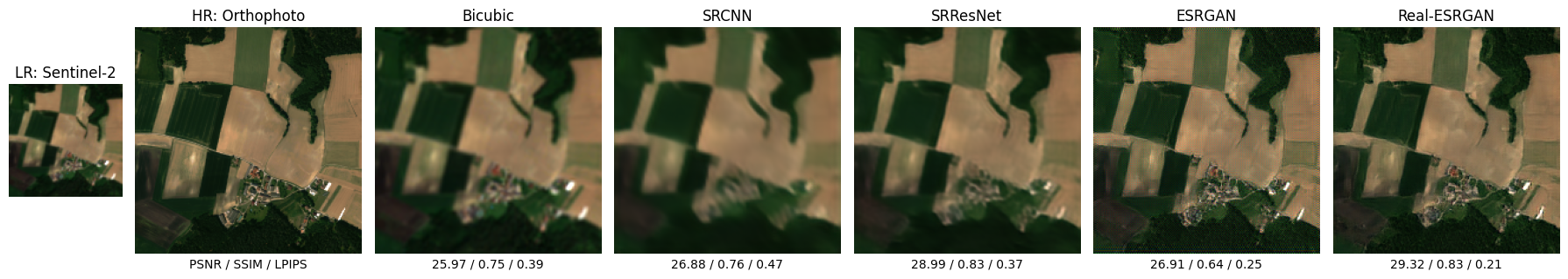}
\end{subfigure}
\caption{Visual comparison of representative patches: Baseline vs. GAN models inclusive PSNR / SSIM / LPIPS metric values.}
\label{fig:visual_comparison}
\end{figure*}

As mentioned in Section~\ref{sec:SRMethods}, the training of the GAN models is divided into two stages. First, the generator is pre-trained with an \(L_{1}\) loss function and an initial learning rate of \(2\times10^{-4}\). The learning rate is halved if there is no improvement in the PSNR on the validation dataset after ten epochs, and training is terminated if there is no improvement in PSNR over 25 epochs.
In the second training phase, the discriminator and generator networks are trained alternately. Here the weights of the loss terms in Equation~\ref{equ:3}) for the generator are set to \(\lambda = 5\times10^{-3}\) and \(\eta = 1\times10^{-2}\), and for the \(L_{\text{percep}}\) summand, the five VGG19 layers are weighted following \cite{Wang2021}, with \{0.1, 0.1, 1, 1, 1\}. The learning rates for the discriminator and generator are set to \(1\times10^{-4}\) and are halved every 500 epochs until the end of the training at 2000 epochs. Optimization is performed using the ADAM optimizer \cite{Kingma2014} with the parameters \(\beta_1 = 0.90\) and \(\beta_2 = 0.99\).
For comparison, all models were applied to the same test dataset and evaluated quantitatively and visually. At this point note that for an objective evaluation, the median and mean of the metrics were calculated and the models compared accordingly. The results are summarized in Table~\ref{tab:results}. For the visual comparison, Figure~\ref{fig:visual_comparison} gives representative results. Detailed inspection of the table and figure allow the conclusion that the baseline models SRCNN and SRResNet, which were optimized on pixel-based metrics, are not suitable for boosting Sentinel-2 resolution. They tend to produce very blurred images, partly due to a georeferential offset between the data sources and differences in the capturing technology, which is also reflected in the LPIPS metric. In contrast, the two GAN architectures generate much better qualitative results, where the Real-ESRGAN even exhibits the best LPIPS values as illustrated in Table~\ref{tab:results}. The adversarial training process and the use of the perceptual loss function allow the models to use more information during the learning phase as the pixel-based approaches in the CNN-frameworks, thereby capturing finer details and textures more effectively. This observation is confirmed by looking at the LPIPS metric values in Table~\ref{tab:results}, which show significantly better values for the GAN models compared to the baseline models.
\begin{table*}[b!]
\centering
\caption{Results for the test dataset}
\label{tab:results}
\renewcommand{\arraystretch}{1.2}
\begin{tabular}{ccccccccccc}
\hline
SR-method $\rightarrow$ & \multicolumn{2}{c}{Bicubic} & \multicolumn{2}{c}{SRCNN} & \multicolumn{2}{c}{SRResNet}& \multicolumn{2}{c}{ESRGAN} & \multicolumn{2}{c}{Real-ESRGAN} \\ 
Metric $\downarrow$    & Mean        & Median    & Mean        & Median      & Mean       & Median       & Mean        & Median       & Mean          & Median          \\ \hline
PSNR      & 24.08           & 23.84             & 25.11          & 24.86            & \textbf{26.41}           & \textbf{26.12}            & 24.33            & 24.21        & 25.91           & 25.81     \\
SSIM      & 0.59           & 0.59             & 0.62          & 0.62            & \textbf{0.68}           & \textbf{0.68}           & 0.52             & 0.53      & 0.64             & 0.64        \\
LPIPS     & 0.46           & 0.46             & 0.56          & 0.57            & 0.48           & 0.49            & 0.31             & 0.31      & \textbf{0.28}             & \textbf{0.28}          \\ \hline
\end{tabular}
\end{table*}
When comparing the GAN models, the Real-ESRGAN achieves better results in all three metrics due to its U-Net discriminator approach. However, evaluating models solely based on PSNR and SSIM reveals that SRResNet outperforms all other models. This is because CNN-based models optimize on a pixel-by-pixel basis, resulting in higher values for these metrics.
Another important aspect that needs to be considered is the significant drop in performance of SR methods compared to the results of previous studies \cite{Liebel2016, Fu2018, Kawulok2019, Galar2019, Pineda2020, SalgueiroRomero2020, Gong2021}. This decrease is primarily due to the fact that these methods are often developed on synthetic datasets. In such datasets, a LR image is generated from a HR image through a degradation process, which is then used to train the methods. In contrast, the methods originally developed using synthetic datasets are tested on a real-world problem in this work.
\section{Conclusion and Future Work}
In this work, SR techniques were investigated to enhance the spatial resolution of the Sentinel-2 RGB bands from \(10 \times 10\) m to \(5 \times 5\) m. Both, conventional CNN-based methods as well as more complex GAN structures were used and compared. The main objective was to evaluate their effectiveness in a real scenario, for which reason a specific dataset of Sentinel-2 images and aerial orthophotos was created under consideration of adequate pre-processing steps. The results reveal that the CNN-based methods are effective in up-scaling and enhancing image details, but produce blurry results as pixel-based loss functions are applied only for this type of models. In contrast, the GAN models, especially Real-ESRGAN, demonstrated superior ability in generating high quality images, which is also due to the LPIPS metric accounting for human perception. As evaluation reveals promising results in terms of feasibility and boosting capability, especially the GAN models are worth being investigated and optimized further in future. The primary focus here will be to increase the up-scaling factor from \(\times2\) to \(\times4\) in order to achieve a spatial resolution of \(2.5 \times 2.5\) m. Furthermore, the authors will include the NIR band, as it may contain additional information not reflected in the RGB bands. To verify whether synthetically boosted satellite images are suitable to improve land use and land cover applications, the results will be tested on a real-world dataset and checked for classification results.
\section*{Acknowledgments}
The authors acknowledge the generous support of the Carinthian Government and the City of Klagenfurt within the innovation center KI4LIFE. We would also like to extend our sincere thanks to Andreas Windisch for his valuable advice.
\newpage

\newpage

\bibliographystyle{unsrtnat}
\bibliography{references.bib}

\end{document}